\documentclass{aa} 
\usepackage{graphicx} 

\def\Msun{M$_\odot$}

\def\0BMV{$(B~-~V)_{\rm 0}$}

\def\simgt{\lower.5ex\hbox{$\; \buildrel > \over \sim \;$}} 
\def\simlt{\lower.5ex\hbox{$\; \buildrel < \over \sim \;$}}
\begin{document} 
%
%
\title{Helium self--enrichment in globular clusters and
the second parameter problem in M3 and M13}
\author{V. Caloi\inst{1} and F. D'Antona\inst{2}} 
\offprints{} \mail{}

\institute{Istituto di Astrofisica Spaziale e Fisica Cosmica, INAF,
 Roma\\
 \email{caloi@rm.iasf.cnr.it}
 \and
   INAF, Osservatorio Astronomico di Roma\\
  \email{dantona@mporzio.astro.it}}

\date{Received ; accepted }
 
\abstract{Inspection of the CM diagrams of globular clusters having
similar heavy element content shows that the luminosity of the red giant
bump relative to the turnoff ($\Delta V_{\rm TO}^{\rm bump}$) differs by
more than 0.1 mag between clusters with different horizontal branch
morphology. Unfortunately, careful consideration of the data leaves us
with only one pair (M3 and M13) of clusters good for a quantitative
discussion. For this pair we consider differences in age and helium
content as possible causes for the difference in $\Delta V_{\rm TO}^{\rm
bump}$, and find more convincing support for the latter. A larger helium
content in M13 stars (Y $\sim$ 0.28 vs. Y $\sim$ 0.24) accounts for
various CM diagram features, such as the difference in the luminosity
level of RR Lyr variables and of the red giant bump with respect to the
turnoff luminosity, and the horizontal branch morphology. This enhanced
helium can be tentatively understood in the framework of
self--enrichment by massive asymptotic giant branch stars in the first
$\sim$ 100 Myr of the cluster life. A modest self--enrichment can be
present also in M3 and can be the reason for the still unexplained
presence of a not negligible number of luminous, Oosterhoff II type RR
Lyr variables. The hypothesis that a larger helium content is the second
parameter for clusters with very blue horizontal branch morphology could
be checked by an accurate set of data for more clusters giving turnoff,
RR Lyrs and bump magnitudes within a unique photometry.
}

\titlerunning{Helium self--enrichment and the case of M3 and M13}
\authorrunning{ } 
\maketitle

\section{Introduction}
The processes that gave origin to globular clusters (GCs) begin to
appear more and more complex, in the light of chemical inhomogeneities
and peculiarities, difference in radial distribution of the various
cluster stellar components, the permanence of the second parameter
mistery, etc. (f.e., Kraft 1994, Gratton et al. 2001, Briley et
al. 2004; Catelan et al. 2001). In preceding papers (D'Antona
et al. 2002, D'Antona \& Caloi 2004) we examined the possibility of
self--enrichment in helium as a consequence of star formation episodes
out of the material ejected by massive asymptotic giant branch (AGB)
stars of the first generation. Such material will be enriched in N,
possibly in Na and Al, but surely enriched in helium, owing to the
second and third phases of dredge--up. The presence of a spread in
helium content may help in explaining features such as blue tails and
bimodality in horizontal branches (HBs) (see the above quoted papers),
the main point being the reduction in the evolving mass, at a given age,
with increasing helium content.

The most obvious candidates to the helium enrichment phenomenon appear
the clusters with an extended HB, populated in regions difficult
to reach with a standard mass loss of about 0.2 \Msun\ and a dispersion
of 0.02 \Msun\ (f.e., Lee et al. 1994, Catelan et al. 1998). Among these
are many of the so called second parameter clusters, which have an
intermediate heavy element content and a HB mainly populated on the blue
side of the RR Lyrae region. We shall look for features which may be
caused by an enhanced helium content, from a morphological and
evolutionary point of view, leaving aside other considerations
(formation, dynamics, ...), since our purpose is to explore the
possibilities opened by the original helium enhancement hypothesis
without the presumption of offering a complete solution to the many
problems involved.

\begin{figure}
\begin{center}
\resizebox{8.8cm}{!}{\rotatebox{0}{\includegraphics{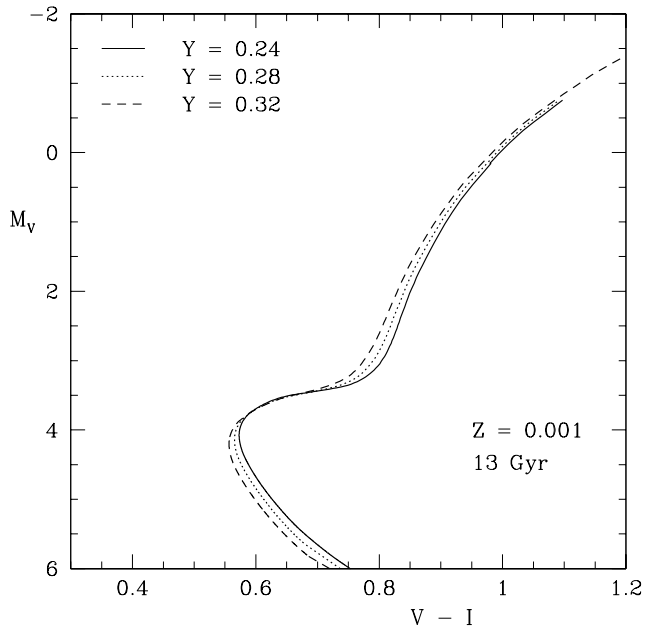}}}
\end{center}
\caption{Isochrones of 13 Gyr for Z=0.001 and Y=0.24,0.28 and 0.32 (see text).
\label{fig1} } \end{figure}

\begin{figure}
\begin{center}
\resizebox{8.8cm}{!}{\rotatebox{0}{\includegraphics{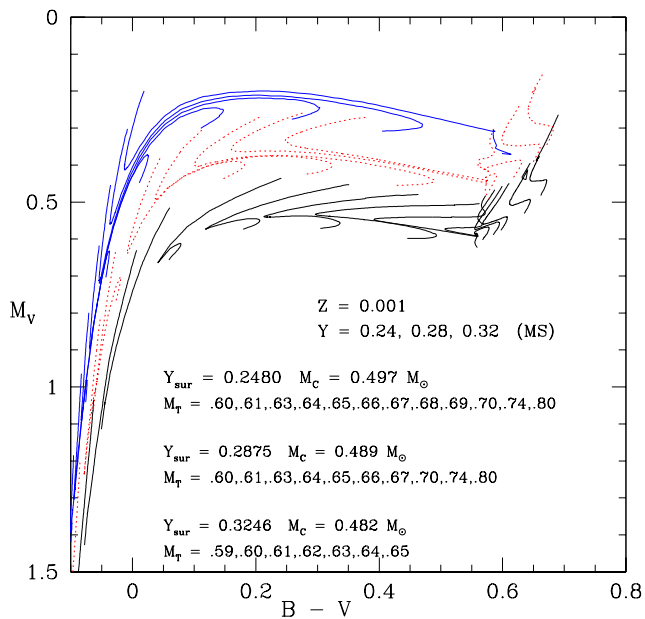}}}
\end{center}
\caption{Theoretical HB evolutionary tracks for Z=0.001 and MS helium
content Y=0.24, 0.28 and 0.32. Total mass (M$_{\rm T}$), helium core
mass (M$_{\rm C}$) and surface helium content (Y$_{\rm sur}$) are
indicated (see text).
\label{fig2} } \end{figure}

\section{The models}
For the purpose of this investigation, we computed isochrones and HB
models, which are shown in Figs. 1 and 2. The code used is ATON2.0 and
its adjournments (Ventura et al. 1998). The heavy element content Z =
0.001 has been chosen as close to the value observed in the intermediate
metal poor GCs that we shall consider, and in particular, close to the
metallicity of M3 and M13 (f.e., Sneden et al. 2004). A helium content
of 0.24 has been assumed, close to the cosmological value (Izotov \&
Thuan 1998, Spergel et al. 2003).

As we see from Fig. 1, the turnoff in isochrones with the same metal
content and increasing helium content becomes bluer and fainter.  The
turnoff at 13 Gyr and Y = 0.28 is 0.06 mag fainter than with Y = 0.24,
and the shift in colour is of $\sim$ 0.007 mag, both in (B--V) and
(V--I). The HB evolutionary tracks with varying Y have been computed
with the evolutionary helium core masses (M$_{\rm c}$ in Fig. 2) and
surface helium content deriving from the first dredge--up (Y$_{\rm sur}$
in Fig. 2), and are terminated when central Y is reduced to 0.10. The
most important feature of the tracks is the increase in luminosity with
increasing Y (Sweigart \& Gross 1976): the HB with Y = 0.28 is on the
average $\sim$ 0.20 mag more luminous than the one with Y = 0.24.

\section{The bump on the red giant branch}
\subsection{Selection of the sample}
The slowing down of red giant (RG) evolution which takes place when the
hydrogen shell reaches the point of the maximum inner expansion of the
convective envelope gives origin to the so called red giant bump (Thomas
1967, Iben 1968); this feature has been observed by now in many clusters
(Ferraro et al. 1999, Zoccali et al. 1999, Riello et. 2003). We consider
intermediate metallicity clusters with very similar heavy element
content, and report in Table 1 the magnitude difference between
turnoff and RG bump.  We considered only clusters for which the data of
the turnoff and bump come from the same photometry.\footnote{For this
reason for NGC 6752 we used the photometry by Buonanno et al. (1986),
the only set of data in which turnoff and RG bump are both
identified.}  Unfortunately, it is not possible to use recent large data
sets: i) Rosenberg et al. (1999) give a homogeneous data set
for turnoffs and HBs, but not for RG bumps; ii) RG bumps magnitudes
are given by Ferraro et al. (1999) (the ones we used), by Zoccali et
al. (1999), based on HST observations transformed into the standard B
and V system, and by Riello et al. (2003), who leave the data in the HST
system. They choose to do so because of
possible ``deceptive errors in the estimate of the visual magnitudes''
introduced by the transformation to the standard Johnson system, which
requires knowledge of the cluster reddening. Both Zoccali et al. (1999)
and Riello et al. (2003) do not give turnoff estimates.

In any case, we used Rosenberg's turnoff magnitudes together with the
ones quoted in present paper to reduce the errors, except for NGC 7089
and NGC 6934, not present in Rosenberg's compilation. A meaningful
improvement is found only for M3 and M13, for which the errors in
Rosenberg's data are substantially lower than those in Table 1. As for
the bump magnitudes, the Riello ones are useless for our purpose, and
the ones by Zoccali et al. may be used for clusters with low reddening,
to reduce the calibration errors mentioned before.  We stress that there
are no other intermediate metallicity clusters in the literature for
which turnoff and bump magnitude are derived from the same photometry.

\subsection{A hint for differences in $\Delta V_{\rm TO}^{\rm bump}$}
In Table 1 we compare the data for clusters with mostly blue
HBs with the data for clusters with HBs well populated in the B, V and R
regions. We found only two clusters of this latter type, one of which is
the well studied M3.\footnote{We did not consider NGC 6229 ([Fe/H] =
-1.30) for which the bump magnitude is uncertain (Borissova et
al. 1997), and NGC 3201 which is 0.1 dex more metal rich than M3 ([Fe/H]
= -1.23, Carretta \& Gratton 1997).}

\begin{table*}
  \caption[]{Globular Clusters with very similar heavy element content
  in Carretta \&Gratton scale
  and differing CM diagram features}
\medskip
   \begin{tabular*}{18cm}{ccccccccc}   \hline \\ 
 Cluster & Messier & [Fe/H]$_{\rm CG}$&  V$_{\rm TO}$  &   V$_{\rm bump}$ & $\Delta V_{\rm TO}^{\rm bump}$ &    V$_{\rm RR}$  &  N$_{\rm RR}$  \\  
\hline 
\noalign{\smallskip}
\multicolumn{4}{c}{}  Globular Clusters with predominantly blue HB && \\ 
 NGC 1904& M79 &-1.37 & 19.70 $\pm$ 0.10 & 15.95 $\pm$ 0.10 & 3.75 $\pm$ 0.11 & & & \\
 NGC 6093& M80 &-1.41 & 19.75 $\pm$ 0.10 & 15.95 $\pm$ 0.10 & 3.80 $\pm$ 0.14 & 15.98   &  3 \\
 NGC 6205& M13 &-1.39 & 18.55 $\pm$ 0.10 & 14.75 $\pm$ 0.05 & 3.80 $\pm$ 0.11 &14.83 $\pm$ 0.02& 7 \\
 NGC 6752&     &-1.42 & 17.40 $\pm$ 0.10 & 13.65 $\pm$ 0.05 & 3.75 $\pm$ 0.11 &         & \\
 NGC 7089& M2  &-1.38 & 19.60 $\pm$ 0.10 & 15.85 $\pm$ 0.05 & 3.75 $\pm$ 0.11 & 15.926 $\pm$ 0.021& 11 \\
\hline 
\noalign{\smallskip}
\multicolumn{5}{c}{} Globular Clusters with HB populated in the B, V and R regions && \\ 
 NGC 5272 & M3&  -1.34& 19.10 $\pm$ 0.10 & 15.45 $\pm$ 0.05 & 3.65 $\pm$ 0.11 & 15.665 $\pm$0.013 &35 \\
 NGC 6934 &   & -1.30 & 20.40 $\pm$ 0.15 & 16.78 $\pm$ 0.10 & 3.62 $\pm$ 0.18 & 16.873 $\pm$0.017 &24  \\  
\hline \\
\end{tabular*}
 Bibl. CM diagram: M79: Kravtsov et al. 1997; M80: Brocato et al. 1998; M13: 
 Paltrinieri et al. 1998; NGC 6752: Buonanno et al. 1986 as in Ferraro
 et al. 1999; M2, M3: Lee \& Carney 1999b; NGC 6934: Piotto et
 al. 1999.

 Bibl. RR Lyr: M80: Wehlau et al. 1990, the 3 RR Lyr with the best photometry; 
 M13: Kopacki et al. 2003; M2: Lee \& 
 Carney 1999a, 11 RRab with stable light curves; M3: Carretta et
 al. 1998, 35 RRab with stable light curves chosen by Lee \& Carney
 1999b; NGC 6934: Kaluzny et al. 2001, 24 RRab with stable light
 curves.
\label{bump}
\end{table*}

We notice that the the average $\Delta V_{\rm TO}^{\rm bump}$ for the
blue HB clusters is always larger than for the M3--type clusters: for
the former ones we have an average of 3.77 $\pm$ 0.07 mag, while for the
latter clusters we have 3.64 $\pm$ 0.11 mag.
To be precise, we should correct this latter value for the difference in
average metallicity between the two groups (--1.39 vs. --1.32), since both
turnoff and bump magnitudes increase with metallicity: this small
correction would move the M3--like average to 3.67 mag. However, we do
not enter in the details of the procedure, because an inspection of the
data in Table \ref{bump} shows that the error on the
difference in $\Delta V_{\rm TO}^{\rm bump}$ between the two groups is
of the same order of magnitude of the difference itself. So we limit the
discussion to the clusters M3 and M13, for which a substantial reduction
of the errors is possible, as mentioned before. In dealing with these
two clusters, we shall follow the universal opinion of considering them
of the same metallicity (see, f.e., Rosenberg et al. 1999).

\subsection{The case of M3 and M13} 
In Table 2 we give the available data for M3 and M13 from Rosenberg et
al. (1999) and Zoccali et al. (1999), the weighted means between these
values and the data in Table 1, and the resulting estimate for
$\Delta V_{\rm TO}^{\rm bump}$. The difference in $\Delta V_{\rm
TO}^{\rm bump}$ between the two clusters is 0.14 $\pm$ 0.09 mag.  Even
if this result is meaningful only at one $\sigma$ level, the fact that
in {\it all} blue clusters for which a consistent measure was possible
$\Delta V_{\rm TO}^{\rm bump}$ turned out larger than in M3--type
clusters, appears intriguing enough to be worth of further
investigation.

\begin{table*}
  \caption[]{Additional data for M3 and M13 from Rosenberg et al. and 
 Zoccali et al., plus the weighted means with relative errors when used
 together with data in Table 1}
\medskip
   \begin{tabular*}{18cm}{ccccccccc}   \hline \\ 
 Cluster & Messier & V$_{\rm TO}$(R)&   V$_{\rm bump}$(Z) &V$_{\rm
 TO}$(mean) &V$_{\rm bump}$(mean) & $\Delta V_{\rm TO}^{\rm bump}$ &  $\Delta V_{\rm TO}^{\rm RR}$  \\  
\hline 
\noalign{\smallskip}
 
 NGC 6205& M13 & 18.50 $\pm$ 0.06 & 14.70 $\pm$ 0.04 & 18.51 $\pm$
 0.05&14.72$\pm$ 0.03 & 3.79 $\pm$ 0.06 & 3.68 $\pm$ 0.05 \\
 NGC 5272& M3  & 19.10 $\pm$ 0.04 &  ---             & 19.10 $\pm$ 0.04
&15.45$\pm$ 0.05  & 3.65 $\pm$ 0.06 & 3.44 $\pm$ 0.04 \\
\hline \\
\end{tabular*} 
\label{m3m13}
\end{table*}

\section{Possible causes of the difference in $\Delta V_{\rm TO}^{\rm
bump}$: age or helium difference?}
The main candidates to an influence on CM diagram features are age and
helium content, since in present case differences in heavy element
content are excluded. Age influences $\Delta V_{\rm TO}^{\rm bump}$
because the turnoff and the bump fade at a different rate.  For the
metallicity of these clusters (Z $\sim$ 0.001) and for ages of 12--15
Gyr, the fading of the turnoff with age is of about 0.075 mag/Gyr
(present models, D'Antona et al. 1997, Cassisi et al. 1999). 
For the bump magnitude we have a fading of 0.035 - 0.04 mag/Gyr
(Ferraro et al. 1999, Cho \& Lee 2002, and our own models). So for each
Gyr of difference, $\Delta V_{\rm TO}^{\rm bump}$ would increase by 
about 0.04 mag. 

The difference of 0.14 $\pm$ 0.09 mag would correspond to an age
difference of 3.5 $\pm$ 2.2 Gyr. A similar estimate of 3 Gyr is
often obtained from the difference in HB morphology for the age interval
12--15 Gyr (see, f.e., the discussion in Johnson \& Bolte 1998). For
lower ages -- M3 of about 10 Gyr -- the age difference becomes of the
order of 2 Gyr (Rey et al. 2001).

Let us now consider the effects of an increase in the helium content in
M13. In clusters with Y $\sim$ 0.28 the turnoff luminosity would be
fainter by about 0.06 mag (our models, see Fig. 1 and D'Antona
et al. 2002), while the luminosity of the RG bump would increase by
$\sim$ 0.08 -- 0.09 mag, according to our models, and of 0.09 mag
according to Riello et al. (2003).\footnote{We assumed $\Delta M_{\rm
bol}\sim 1.21\Delta M_{\rm V}$ as in Cho \& Lee 2002.} So $\Delta V_{\rm
TO}^{\rm bump}$ = 0.14 $\pm$ 0.09 would be given by a helium variation
$\Delta$Y = 0.04 + 0.02 / - 0.03.

\subsection{The role of RR Lyrae variables}
We consider now the characteristics of the RR Lyrae variables in
the two clusters. M3 has a population of almost 200 RR Lyr variables,
with a large majority of RRab--type; the average periods are of 0.56 d
and of 0.32 d, for RRab and RRc, respectively (see, e.g., Jurcsik et
al. 2003, Clement et al. 2001). In M13 only 9 RR Lyr stars have been
identified, out of which only one is an RRab; the average period of 7
well observed RRc stars is 0.36 d (Kopacki et al. 2003). Both the
average periods and the percentage of RRc stars indicate that variables
in M3 belong to the Oosterhoff type I (a well established fact), while
those in M13 more likely belong to the Oosterhoff type II. 

Carretta et al. (1998) and Lee \& Carney (1999a,b) tie their RR Lyr
photometries for M3 to the one of the non variable stars (Ferraro et
al. 1997a), obtaining the same average magnitude of 15.66 mag. The same
value is found independently by Jurcsik et al. (2003). So for this
cluster we have a unique photometry for all the data in Table 1.

For M13 the photometry of RR Lyr stars is different from the one of the
CM diagram. The RR Lyr magnitudes are given by Kopacki et al. (2003),
who give also mean values for some bright red giant variables. For these
latter stars, the average of their mean magnitudes differs by $\sim$
0.04 mag from the average obtained from the photometry given by
Pilachowski et al. (1996) for the same objects.\footnote{The variables
V11, V15, V17, V18, V19 and V24 have been considered.} This indicates
that the two photometries should be consistent within 0.04 magnitudes.

The difference in Oosterhoff type between M3 and M13 is part of the well
known second parameter problem: GCs with the same heavy element content
and very different HB populations. From Table 2 we see that
the luminosity difference between turnoff and RR Lyraes $\Delta V_{\rm
TO}^{\rm RR}$ is much larger in M13 than in M3: 3.68 $\pm$ 0.05 vs. 3.44
$\pm$ 0.04 mag (Table 2), so that the difference in
$\Delta V_{\rm TO}^{\rm RR}$ between the two clusters is 0.24 $\pm$ 0.06
mag. In the literature, such a difference is generally interpreted as
due to a higher luminosity of the blue and RR Lyr regions in M13; these
stars are supposed to be in a more advanced, and more luminous,
evolutionary phase than the ZAHB and its vicinity, where most of RR
Lyraes in M3 are found. This behaviour would follow from the higher age
of M13: the smaller HB masses would populate the ZAHB only on the blue
side of the RR Lyr variables.

\subsection{The case of a difference in age}

If the difference of 0.14 $\pm$ 0.09 mag between $\Delta V_{\rm TO}^{\rm
bump}$ in M3 and M13 (Table 2) is all due to an age difference, M13
should be older by 3.5 $\pm 2.2$ Gyr. In this case, the turnoff in M13
should be fainter than in M3 by $\sim 0.075$ mag/Gyr $\times 3.5$ Gyr =
0.26 mag. However, available observations are in strong
disagreement with such a large difference in turnoff luminosity (e.g.,
Ferraro et al. 1997b, Johnson \& Bolte 1998, Rey et al. 2001). Besides, the RR
Lyr luminosity level in M13, following the observed  $\Delta V_{\rm TO}^{\rm
RR}$ (Table 2) would turn out 0.02 mag fainter than in M3, and
the apparent Oo II type of the RR Lyr variables in M13 would remain 
unexplained.

Only considering the {\it minimum} difference in age required by the
observed difference in $\Delta V_{\rm TO}^{\rm bump}$ taking into
account the estimated errors (see above), that is 1.3 Gyr, 
we obtain a consistent
scenario: the turnoff in M13 would be fainter by 0.10 mag than the
turnoff in M3, and the RR Lyrs in M13 would be 0.14 mag brighter than in
M3.  Still, we must remember that such a small difference in age could
lead to the observed difference in HB morphology only for very low ages,
$\sim 10$ Gyr for M3 (Rey et al. 2001).

\subsection{The case of a difference in helium}

If the age is the same and helium in M13 is enhanced up to Y = 0.28, we
have that the turnoff in M13 would be fainter by 0.06 mag and the HB
would be brighter by 0.20 mag (see Fig. 1 and
Fig. 2). This explanation is consistent with the observed
difference in $\Delta V_{\rm TO}^{\rm RR}$ between M13 and M3 (0.24
$\pm 0.06$ mag). In addition, it predicts for the RR Lyraes in M13 a larger
luminosity (by $\sim$ 0.20 mag) than for those in M3.\footnote{To be
precise, the RR Lyrs become brighter by 0.18 mag.}

As a conclusion, the hypothesis of a difference in the helium content of
the stars in these two clusters is the more appealing one, if the
$\Delta V_{\rm TO}^{\rm bump}$ will be confirmed to the level used in
this work. Therefore, we stress that it would be extremely important to
have homogenous sets of data for the turnoff {\it and} bump. In fact, we
could carry on this kind of analysis only for the famous couple M3 -- M13.

\section{The detailed fitting of M3 and M13 with differing helium
contents} 

Beside the relative luminosity levels, we have to check whether the
(subtle) differences in the CM diagram introduced by a helium increase
find some support in the observations.  In Fig. 3a it is shown the fit
of the fiducial lines for M3 given by Johnson \& Bolte (1998) with the
isochrone of 13 Gyr, Z=0.001, Y=0.24, and the relative HB
models. Similarly, in Fig. 3b the fit is shown for M13, but with an
isochrone corresponding to Y = 0.28; the average position of six RR Lyr
variables from Kopacki et al. (2003) is also indicated. Different values
for Y not only provide a better explanation for the relative
luminosities of the turnoff, HB and bump, but also a better
morphological fit. To further stress this point, we show in
Fig. 3c a fit of M13 fiducial sequence with isochrones with
Y=0.24, ages of 13 and 14 Gyr: there is an evident difficulty in
obtaining a satisfactory fit, in contrast with the increased helium
case.

\begin{figure}
\begin{center}
\resizebox{6.8cm}{!}{\rotatebox{0}{\includegraphics{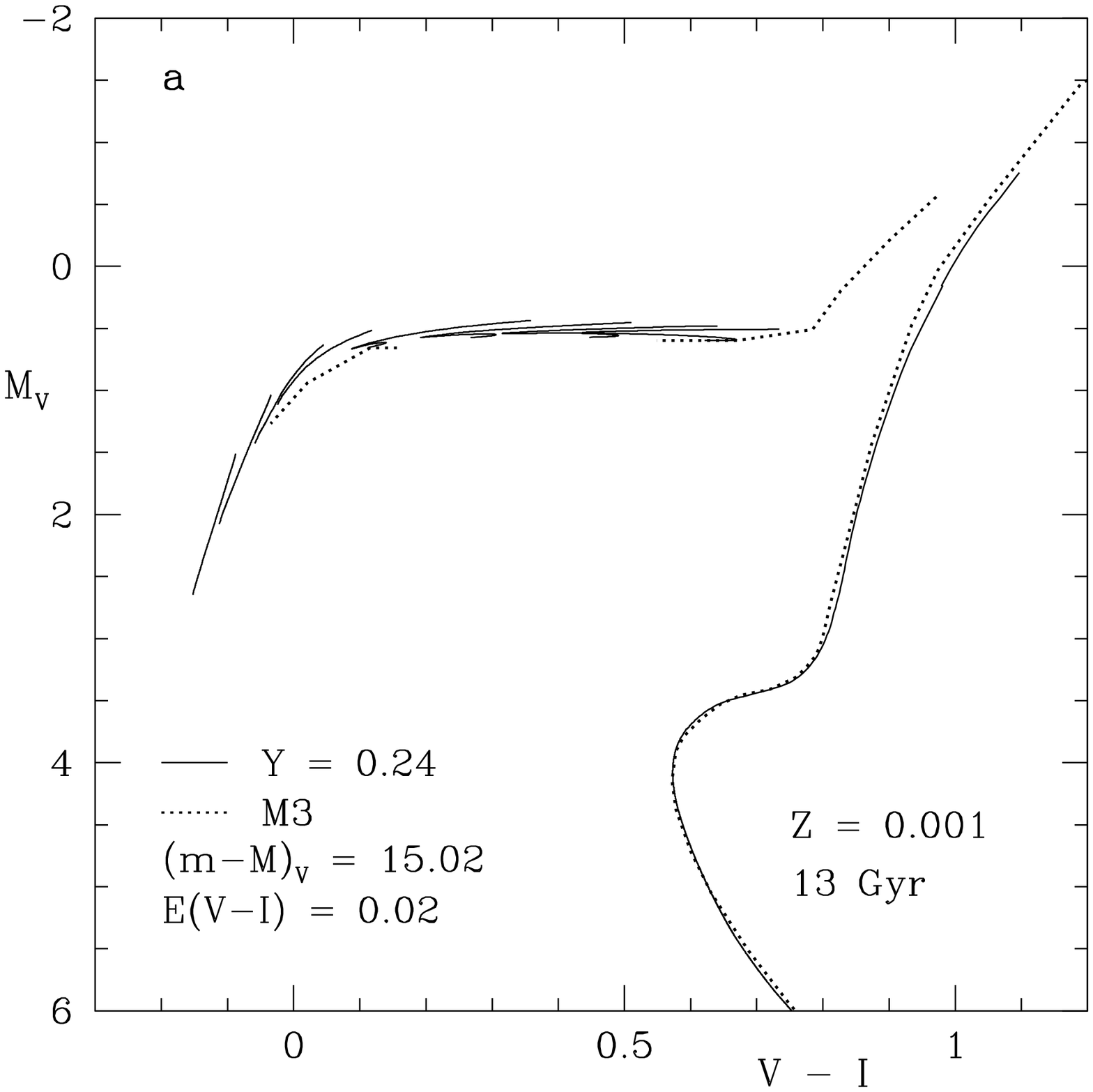}}}
\resizebox{6.8cm}{!}{\rotatebox{0}{\includegraphics{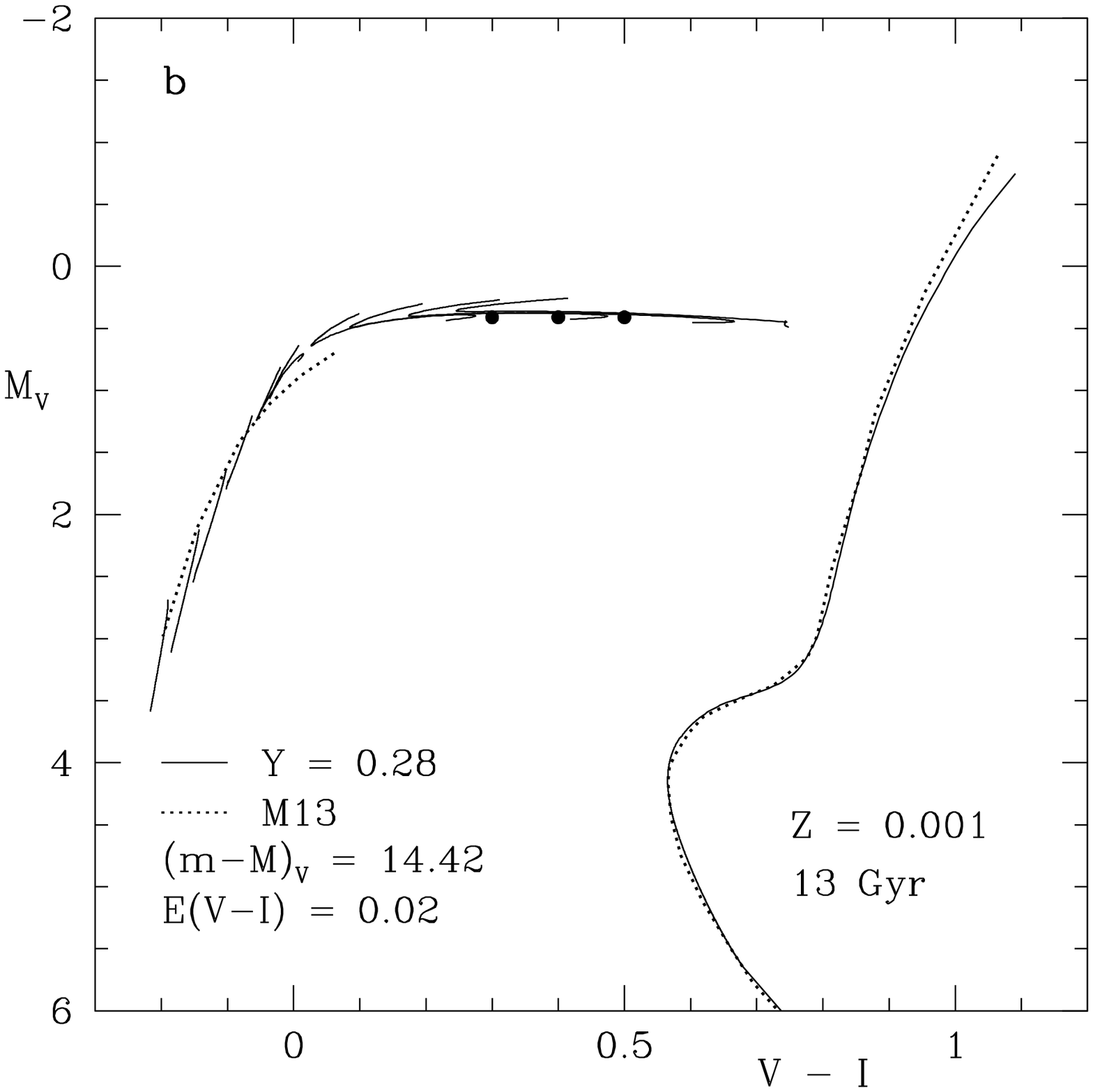}}}
\resizebox{6.8cm}{!}{\rotatebox{0}{\includegraphics{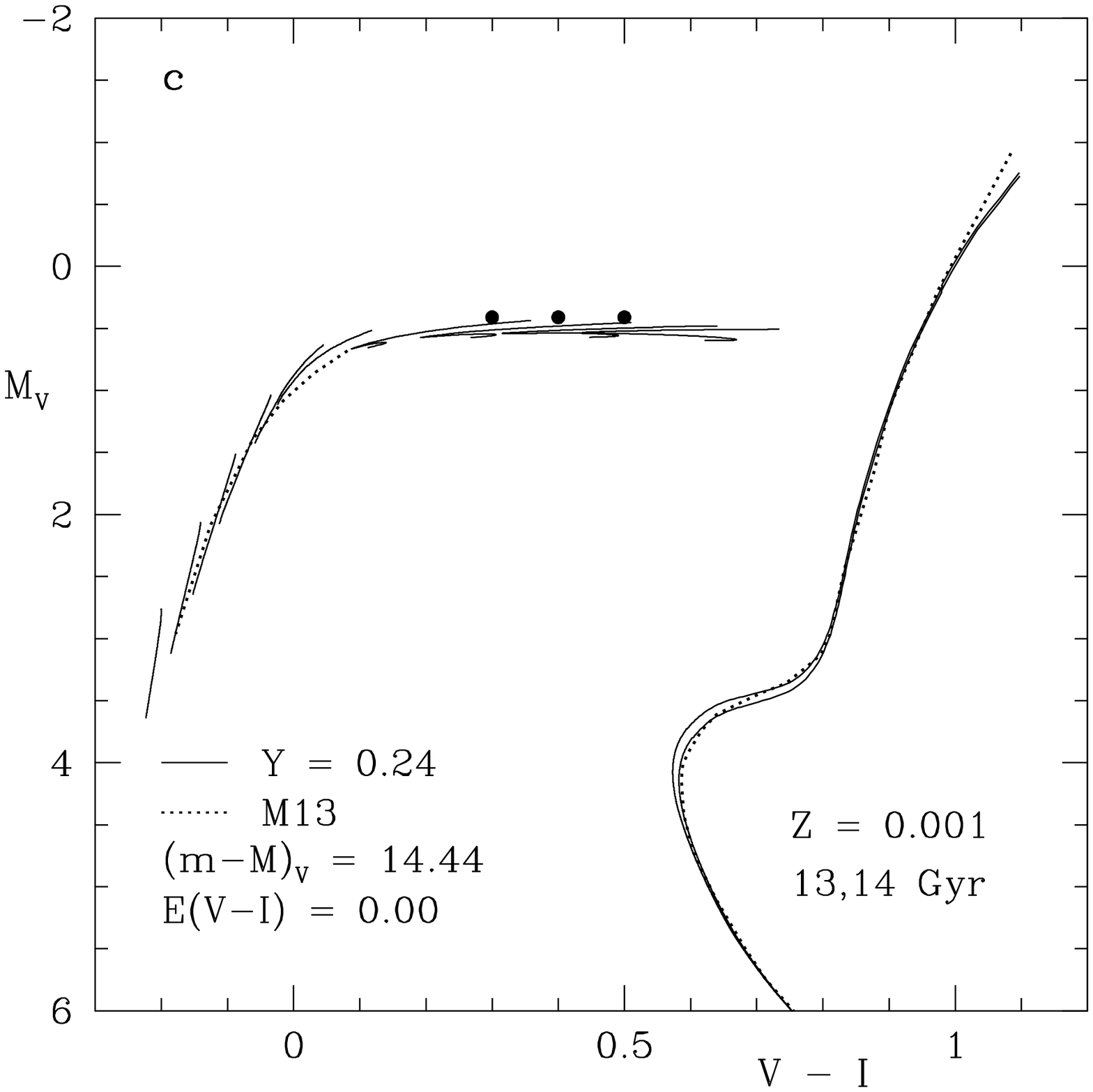}}}
\end{center}
\caption{Fig. 3a: CM diagram of M3 (data by Johnson \& Bolte) compared
with an isochrone of 13 Gyr, Y=0.24 and relative HB tracks. Fig. 3b: CM
diagram of M13 (data by Johnson \& Bolte, RR Lyr level from Kopacki et
al., full dots) compared with an isochrone of 13 Gyr, Y=0.28 and
relative HB tracks. Fig 3c: CM diagram of M13 (data by Johnson \& Bolte)
compared with isochrones of 13 and 14 Gyr and relative HB
tracks for Y=0.24. The Y=0.28 models give a better morphological fit of the
turnoff region of M13.
\label{fig3} } 
\end{figure}

Finally, the HB stellar distribution (second parameter problem) finds a
natural explanation. In D'Antona et al. (2002) we examined the
consequences on HB population of an increase in Y from 0.24 to 0.28. To
complete the discussion, we summarize the main effects. In M3, assuming
Y = 0.24 on the main sequence, we have that the bulk of the HB
population covers the mass interval 0.60--0.70 \Msun, with the
peak at the RR Lyr region; a sparse tail toward the blue would require
lower masses. Since the evolving giant (in absence of mass loss) at the
helium flash is of about 0.82 \Msun, the implied mass loss is 0.17 $\pm$
0.05 \Msun.

With a helium content of 0.28 and 13 Gyr of age, the evolving giant at
the helium flash is of 0.766 \Msun (Z = 0.001, as before). With the same
mass loss as in M3, the maximum HB mass turns out 0.646 \Msun, in the
middle of the RR Lyr region (see Fig. 2), while the minimum
mass is of about 0.546, almost down at the level of the turnoff. If mass
loss along the giant branch increases with decreasing mass (e.g., Lee et al.
1994), lower masses can be reached, but see also Lee \&
Carney (1999b). So M13 would naturally shift its HB population
according to observations. Let us note that we have been mentioning a
helium content of 0.28 for simplicity, while, most likely, a spread in Y
is to be presumed, starting from $\sim$ 0.28 and reaching well beyond
0.30 (D'Antona \& Caloi 2004). We elaborate briefly on the subject of
self--enrichment in the following section.

\section{The self--enrichment scenario}

In a preceding paper (D'Antona et al. 2002) we considered helium
enrichment as mostly related to the origin of very blue HB tails in
second parameter clusters such as M13 and NGC 6752.  We are now
proposing that the overall difference between M3 and M13 CM diagram
morphologies is due to the different helium content. Thus {\it all}
stars in M13 should have an enriched helium content with respect to the
Big Bang abundance. How is this possible? We suggest that M13 represents
an extreme case of self--enrichment. Among GCs which show abundance
spreads, we have identified NGC 2808 as a cluster in which at present
half of the stellar population has a normal Y, and the other half is
helium enriched at various degrees (D'Antona \& Caloi 2004). The normal
Y population is responsible for the red part of the HB, and the enriched one
is responsible for the blue side of the HB and for the blue tails. We
have shown that this
may happen only if the number of low mass stars with normal Y content
present in the cluster is much smaller than expected on the basis of an
initial mass function which is, on the one side not too implausible, and
on the other, able to explain the blue HB stars as born from the helium
enriched ejecta of the massive AGBs. Then many low mass stars of the
first generation must have been lost by the cluster, perhaps because the
intermediate mass stars were much more concentrated into the central
regions (D'Antona \& Caloi 2004).  In M13 {\it there is no red HB
clump}: in this case, the first generation stars must have been
completely lost by the cluster.
This hypothesis is certainly difficult to be digested without a detailed
modeling: in fact M3 and M13 are very similar on any other respect:
present day mass, central density, relaxation time. However, this kind
of scenario is still at the beginning. Notice in addition, that Salaris
et al. (2004), in a detailed analysis of the R parameter in GCs,
detect a significant spread of helium content 
towards higher abundances, in the clusters having very blue
HBs. This is a further indication in favour of our hypothesis.

Beside the difficulties with the cluster dynamics, there is the other
one that, in this scheme, {\it all} M13 stars should be helium enriched,
while the observations show that not all of them are oxygen poor (f.e.,
Sneden et al. 2004), a feature which is considered to accompany
processed matter. However, it has been shown that, while most of the AGB
matter from which the second generation stars are born is helium
enriched due to the action of the second dredge--up, not necessarily all
this same matter is oxygen depleted (Ventura et al. 2001, 2002).

Is then M3 devoid of self--enrichment? Actually this may be not the case,
as suggested by the presence of luminous RR Lyrae variables.

\subsection{The case of the luminous RR Lyrae variables in M3}
Recently it has been noticed that a few RR Lyr variables in M3, the
prototype of the Oo I type clusters, show Oo II characteristics (Clement
\& Sheldon 1999, Corwin \& Carney 2001, who quote similar observations
by Belserene 1954). Jurcsik et al. (2003) made a detailed study of about
150 RR Lyr stars (of both types ab and c), and found that they could be
classified in four groups, according to their mean magnitudes and
periods. The most luminous sample has statistically Oo II properties
regarding the mean periods and RR$_{\rm ab}$/RR$_{\rm c}$ number ratio.  
They found also that the various samples can be identified with
different stages of HB evolution, with some difficulty for the most
luminous one. In fact, this has a luminosity at which evolution
is rather fast (last phases of central helium burning), too fast to
support the existence of more than 20\% of the observed RR Lyr stars. 
As Jurcsik et al. (2003) remark, the discrepancy could be removed by the
presence of the (infamous)``breathing pulses'', that is, of a final
phase of helium mixing in convective cores, when helium abundance is
approaching zero (f.e., Castellani et al. 1985, Dorman \& Rood 1993,
Caloi \& Mazzitelli 1993), but at present the phenomenon is not
considered real and HB evolution is computed ignoring it.

Keeping in mind such a possibility, we can anyway consider other
solutions to the problem of Oo II RR Lyrs in M3. On the line of multiple
star generations with helium enrichment, we can hypothise that also this
cluster presents a certain amount of second generation members formed
from the helium rich ejecta of AGB stars. Beside the luminous RR Lyrs,
other features could be interpreted in terms of variable helium content:
i) the large range in period covered by RR$_{\rm d}$ variables
(Clementini et al. 2004); ii) the great length of the HB, since the
lower mass of helium rich red giants, at a given age, helps to originate
the very blue HB stars hotter than the main body of HB population
(Ferraro et al. 1997a, Fig.  18); the spread in luminosity and colour in
the subgiant region (as noted by Clementini et al., see Fig. 4 in Corwin
\& Carney 2001, and Fig. 15 in Ferraro et al. 1997a).
In this respect, we notice that, according to recent estimates (Sneden
et al. 2004), about one--third of the observed giants in M3 are
oxygen poor. In the hypothesis mentioned before (conversion of oxygen to
nitrogen in the envelopes of AGB stars of a preceding generation),
these stars could give origin to the luminous RR Lyr variables, being
also helium enriched.

\section{Discussion}
The recently acquired impressive amount of detailed information on the
chemical composition of GC members down to the main sequence has
profoundly changed our perception of these most ancient stellar systems.
Complex processes, both chemical and dynamical, must have taken place in
their formation phase, leaving characteristic marks on stellar chemistry
and population. 

Here we have tried to relate some photometric features in CM diagrams to
chemical peculiarities, such as helium enrichment. Johnson \& Bolte
(1998) suggested a higher helium content in M13 as the best explanation
for the CM diagram morphology differences between this cluster and M3,
and discussed widely the subject. Beside the turnoff shape (see
Fig. 3), we consider the luminosity
of the RG bump with respect to the turnoff, and find that it correlates
with the HB type: the metal content being the same, clusters with a blue
HB have a larger $\Delta V_{\rm TO}^{\rm bump}$ than clusters with a
uniformly populated HB.

Considering in detail M3 and M13, we examine both the difference in age
and in helium content as possible causes for this feature, and find that
the latter is the more plausible, if one considers the luminosity level
of the RR Lyr variables. The presence of Oo II type RR Lyrs in M3 can
be also interpreted in terms of a helium enriched stellar component.

The difference in helium content between M3 and M13 appears as a
possible ``second parameter'', at least with regard to this pair of
clusters. The difficulties are various, mainly deriving from our
ignorance of GC formation phases and of the initial mass
function(s). Anyway, with respect to similar suggestions in the past,
the hypothesis presented here has the support of detailed investigations
on the AGB ejecta composition and of the recognized necessity of a
primordial contamination of main sequence stars in many clusters, beside
the interpretation of CM features otherwise not easily understood (HB
and bump luminosities). Partial pollution does not help much to give
origin to very blue HB structures (Caloi 2001, D'Antona et al. 2002),
while an increased structural helium content does (D'Antona et
al. 2002). An important contribution to the problem would be to check
the presence of chemical peculiarities in the luminous, OoII type RR Lyr
variables in M3, that we proposed as candidates for second generation, helium
increased structures.

\end{document}